\documentclass[pra,aps,twocolumn,10pt,showpacs,nofootinbib]{revtex4-2}
\usepackage[utf8]{inputenc}
\usepackage[T1]{fontenc}
\usepackage{amssymb}
\usepackage{amsmath,amsthm,mathptmx}
\usepackage{amsfonts}
\usepackage{mathcmd}
\usepackage{graphicx}

\usepackage{braket}
\usepackage{bm}
\usepackage{xcolor}

\usepackage{tikz}
\usepackage{rotating}
\usepackage{float}

\usepackage[colorlinks=true, linkcolor=purple, citecolor=blue]{hyperref}

\begin{document}

\title{Non-Markovian interactions with pulses of quantum radiation}

\author{Julien Pinske}
\email{j.pinske@uni-rostock.de}
\affiliation{Institut für Physik, Universität Rostock, Albert-Einstein-Straße 23-24, D-18059 Rostock, Germany}

\author{Stefan Scheel}
\affiliation{Institut für Physik, Universität Rostock, Albert-Einstein-Straße 23-24, D-18059 Rostock, Germany}

\date{\today}

    \begin{abstract}
        The interaction of a traveling pulse of quantum light with a localized quantum system shows non-Markovian features when scattering takes place in a structured environment.
        Here, we devise a non-Markovian input-output theory by coherently coupling the scatterer to a pseudomode, which decays into a Markovian reservoir.
        This incorporates memory effects of a structured environment without altering the cascaded nature of the Lindblad master equation, whose solution provides the quantum state of the output field of any desired mode.
        We apply our theory to the stimulated emission by a two-level atom, and to transmission of a Gaussian pulse through a cavity.
        We observe that the non-Markovian revival of coherence in the scatterer distorts the single-mode nature of the incoming pulse, thus resulting in a multimode output field.
    \end{abstract}

    \maketitle

    \section{Introduction}
    \label{sec:intro}

    In the classical theory of light, the interaction of electromagnetic waves with matter is governed by Maxwell's equations.
    There, a localized object imposes boundary conditions on an incoming light field, resulting in a scattering problem whose central task is to determine how the field is transformed by the scatterer.
    Although Maxwell's equations retain their formal appearance under quantization of the electromagnetic field, with Hilbert-space operators replacing the classical fields \cite{SS08}, the description of a traveling pulse of quantum radiation poses a formidable challenge.
    Rather than a single classical wave, the field consists of a continuum of quantized modes, and the quantum state of a finite pulse of light generally occupies a coherent superposition of infinitely many of them.
    As a consequence, exact solutions to the scattering problem are restricted to one- and two-photon states \cite{SF07,F18},
    and numerical solutions become prohibitive beyond the few-photon regime \cite{SS15,PBK17,FTR18,SCC15}.
    Here, quantum optics shares many of the conceptual and computational challenges familiar from quantum field theory \cite{LSZ55}.

    For this reason, input-output theory for quantized light has traditionally been formulated in the Heisenberg picture \cite{CG84,GC85}.
    Alternatively, one expresses field correlation functions in terms of the evolution of the scatterer, leading to the so-called source master equation \cite{G93,C93}.
    Although in this approach the state of the light field is never represented explicitly, many of its nonclassical properties can be inferred from system correlation functions \cite{BT56,G63,SV06,BCB12,HHD26}.
    An important simplification occurs in cavity quantum electrodynamics, where the standing electromagnetic wave inside a cavity is a single-mode field \cite{BMB04,BBM05,KKW05}.
    The coupled atom-field dynamics are described by the Jaynes-Cummings model \cite{JC63}, and, in particular, no multimode scattering occurs \cite{S19,CLY24}.
    More recently, it was recognized that the interaction of a traveling quantum pulse with a localized scatterer is formally equivalent to a situation in which the pulse has been released by a (far-)distant cavity \cite{KM19}.
    Putting this virtual cavity before the scatterer yields a simple cascaded master equation \cite{CKS17} for the scattering problem.
    This tames the multimode character of the incoming light field and provides the quantum state of any outgoing mode \cite{KM20}.

    The temporal multimode structure of quantum light is central to a wide range of applications, including photonic quantum computing \cite{SGS23,BAS26}, quantum communication with flying qubits \cite{GEP98,K08,RR15}, and quantum-enhanced sensing \cite{KM23}.
    On the one hand, the generation of useful nonclassical states of light may require a single-mode output state \cite{HWD19,GLY19}.
    On the other hand, photonic quantum gates have been proposed that make explicit use of the multimode nature of light.
    These include the splitting of a two-photon pulse \cite{LYM23} and photon-number sorting \cite{YLP22}.

    A key assumption underlying the input-output theory for traveling pulses \cite{KM19} is the Markov approximation \cite{BP10}.
    The scatterer is assumed to emit into a vast unstructured continuum that irreversibly removes excitations from the system.
    Many realistic environments, however, possess a nontrivial spectral structure \cite{KB99,SKW99,SCM26}.
    That is, the spectral density varies appreciably over the linewidth of the emitter.
    As a consequence, the scatterer exhibits revivals of coherence and information backflow from the environment, being defining features of non-Markovianity \cite{BLP02}.

\begin{figure}[t]
        \centering
        \begin{tikzpicture}
        \node at (0,0) {\includegraphics[width=0.48\textwidth]{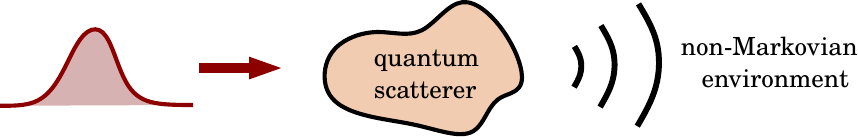}};
        \node at (-4,1.2) {(a)};
        \node at (-2.8,0.7) {$u(t)$};
        \node at (-0.2,1) {$(\hat{H}_s,\hat{c})$};
        \node at (0,-3) {\includegraphics[width=0.48\textwidth]{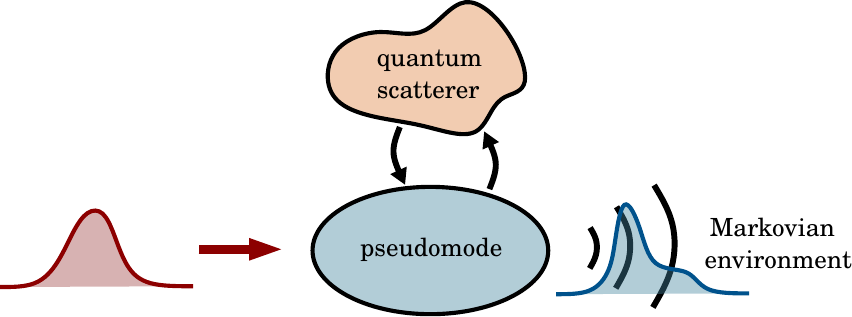}};
        \node at (-4,-1.8) {(b)};
        \node at (-2.8,-3.3) {$u(t)$};
        \node at (0.1,-4.28) {$\hat{a}_p$};
        \node at (-0.7,-3.0) {$\hat{H}_{sp}$};
        \node at (1.8,-3.2) {$v(t)$};
        \end{tikzpicture}
        \caption{\label{fig:synopsis} (a) A quantum pulse $u(t)$ scatters off a quantum system $(\hat{H}_s,\hat{c})$ into a non-Markovian environment.
        (b) The pulse interacts with one (or multiple) pseudomodes $\hat{a}_p$ which coherently exchange energy with the scatterer via a Hamiltonian $\hat{H}_{sp}$, and decays into a Markovian environment, which occupies one or several modes $v(t)$.}
    \end{figure}
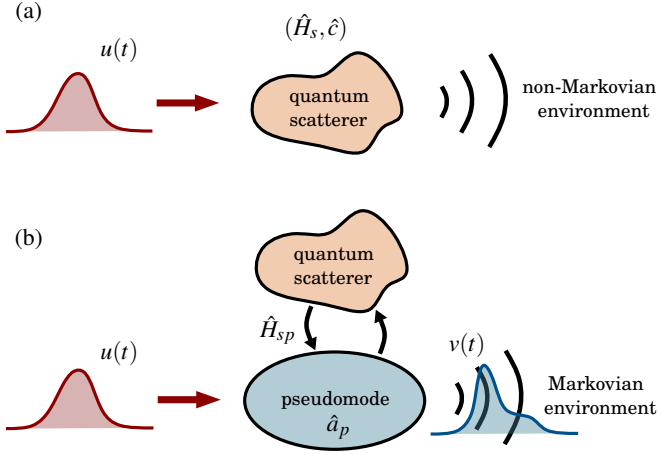

    In this article, we develop a theory describing the interaction of a traveling quantum pulse with a quantum system that emits into a structured, non-Markovian environment; see Fig.~\ref{fig:synopsis}~(a).
    We do so by replacing the non-Markovian dynamics with a unitary coupling to a pseudomode \cite{G97}, which itself decays into a Markovian reservoir; see Fig.~\ref{fig:synopsis}~(b).
    The pseudomode approach yields a master equation in Lindblad form \cite{L76,GKS76} that reproduces the exact non-Markovian dynamics of the scatterer.
    Master equations derived from pseudomodes are therefore more easily solved than other exact techniques, such as the Nakajima-Zwanzig equation \cite{N58,Z60} or the time-convolutionless master equation \cite{CS79,BKP01}, which must often be treated perturbatively through a cumulant expansion \cite{K62,K74,BKP99}.
    The theory of pseudomodes was originally derived for atoms interacting with a single photon \cite{G97,MMP09}.
    Since then, pseudomodes have been generalized to account for multiple excitations \cite{DBG01,DG03}, the interaction with a Gaussian environment \cite{TSH18,MSS20}, and non-Hermitian couplings between the system and the auxiliary modes \cite{PGP20}.
    Recently, it was also shown how pseudomodes can be used to calculate correlation functions of the environment \cite{MGF21,LLL26}.

    Each pseudomode is typically associated with its own loss channel \cite{MMP09}, which serves to reproduce the exact non-Markovian dynamics of the reduced system rather than the state of the output field \cite{LAC19}.
    We circumvent this issue by treating the pulse and the scatterer jointly, so that the pseudomodes are embedded in a cascaded chain, which provides the quantum state of the scattered light.
    The theory is applied to the stimulated emission from a two-level atom driven by a single-photon pulse, and to the transmission of a coherent state through an empty cavity.
    In both cases, we observe that the non-Markovian character of the evolution increasingly distorts the single-mode nature of the incoming pulse, resulting in a multimode output field.

    \section{Interaction with quantum pulses}
    \label{sec:theory}

    Consider a localized quantum system that emits radiation via a coupling operator $\hat{c}$ into the surrounding electromagnetic field.
    The system couples to each frequency $\omega$ of the electromagnetic field according to the Hamiltonian $\hat{H}_{I} = i\hbar\int d\omega\sqrt{J(\omega)}e^{i\varphi(\omega)}\hat{c}\hat{b}^\dag(\omega) - \mathrm{H.c.}$, where $J(\omega)$ is the spectral density and $\varphi(\omega)$ is a relative phase
    (see App.~\ref{app:lin-coup} for details on the model).
    For an unstructured environment one may employ the Markov approximation \cite{BP10}, in which the system is assumed to couple equally to each $\omega$, i.e., $J(\omega)=\gamma/2\pi$ is flat.
    In general, however, the interaction cannot be assumed to be Markovian.
    Instead, many realistic systems have a Lorentzian spectral density \cite{XYF10,XL18,QG26},
    \begin{equation}
        \label{eq:Lorentz}
        J(\omega) =  \frac{g_p^2}{2\pi} \frac{\gamma_p}{(\omega - \omega_p)^2 + \big(\frac{\gamma_p}{2}\big)^2},
    \end{equation}
    centered at $\omega_p$ and of width $\gamma_p$.

    \subsection{Scattering by a quantum pulse}
    \label{ssec:pulse}

    To alleviate the problem of the propagation of quantum light, we consider a pulse $u(t)$ released from a distant (virtual) cavity \cite{KM19}.
    If this cavity is coupled to an input field $\hat{b}_{\mathrm{in}}(t)$ via $\hat{H}_u = i\hbar(g_u^*\hat{a}_u\hat{b}_{\mathrm{in}}^\dag - g_u\hat{a}_u^\dag\hat{b}_{\mathrm{in}})$, the Heisenberg equation for the field operator reads \cite{KM20}
    \begin{equation}
        \label{eq:eom-u}
        \dot{\hat{a}}_u = - \frac{|g_u|^2}{2}\hat{a}_u - g_u \hat{b}_{\mathrm{in}},
    \end{equation}
    with a time-dependent coupling
    \begin{equation}
        \label{eq:out-coup}
        g_u(t)=\frac{u(t)^*}{\sqrt{1-\int_{0}^t|u(t^\prime)|^2 dt^\prime}}.
    \end{equation}
    The specific form of the coupling~\eqref{eq:out-coup} ensures that the released pulse has the desired temporal profile $u(t)$.
    This is formally equivalent to the scatterer being hit by a traveling pulse $u(t)$.
    The Heisenberg equation for the system operator $\hat{c}$ interacting with this pulse is
    \begin{equation}
        \label{eq:eom-sys}
        \begin{split}
            \dot{\hat{c}} & = \frac{i}{\hbar}[\hat{H}_s,\hat{c}] + [\hat{c}^\dag,\hat{c}]\bigg(\hat{f}_{\mathrm{in}}(t) + \int_0^t C(t-t^\prime)\hat{c}(t^\prime)dt^\prime\\
            & \quad + \int_0^t K(t-t^\prime)g_u^*(t^\prime)\hat{a}_u(t^\prime)dt^\prime\bigg),\\
        \end{split}
    \end{equation}
    which is derived in App.~\ref{app:Langevin}.
    Here, $\hat{H}_s$ is the system's internal Hamiltonian, and we defined the memory kernels
    \begin{equation}
        \label{eq:bath-corr}
        \begin{split}
            C(t-t^\prime) & = \int J(\omega) e^{-i\omega(t-t^\prime)} d\omega,\\
            K(t-t^\prime) & = \int \sqrt{\tfrac{J(\omega)}{2\pi}} e^{-i(\varphi(\omega)+\omega (t-t^\prime))}d\omega,\\
        \end{split}
    \end{equation}
    as well as the colored input field
    \begin{equation}
    \label{eq:fin}
    \hat{f}_{\mathrm{in}}(t) = \int_0^t K(t-t^\prime)
        \hat{b}_{\mathrm{in}}(t^\prime)dt^\prime ,
    \end{equation}
    i.e., $[\hat{f}_{\mathrm{in}}(t),\hat{f}_{\mathrm{in}}^\dag(t^\prime)]=C(t-t^\prime)$.
    The kernel $C$ governs the emission of the scatterer, whereas $K$ describes the drive exerted by the pulse.
    When evaluating $K(t-t^\prime)$, integrating the square root $\sqrt{J(\omega)}$ can pose a formidable challenge, which we deal with in App.~\ref{app:sum-Lorentz}.

    \subsection{Pseudomode approach to non-Markovianity}
    \label{ssec:pseudomode}

    For the Lorentzian spectral density~\eqref{eq:Lorentz}, the non-Markovian dynamics~\eqref{eq:eom-sys} are equivalent to the system exchanging energy with a pseudomode $\hat{a}_p$, which itself decays into a Markovian reservoir at rate $\gamma_p$ \cite{G97}.
    When the pulse $u$ interacts with the system, the state $\hat{\rho}(t)$ of the joint light-matter system obeys
    \begin{equation}
        \label{eq:LMEq-pulse}
        \dot{\hat{\rho}} = \frac{i}{\hbar}[\hat{\rho},\hat{H}_{sp}+\hat{H}_{up}] + \hat{L}_u \hat{\rho} \hat{L}_u^\dag - \tfrac{1}{2}\{\hat{L}_u^\dag \hat{L}_u,\hat{\rho}\},
    \end{equation}
    which is derived in App.~\ref{app:pseudo-Lindblad}.
    Here,
    \begin{equation}
        \label{eq:Ham-sys-pseudo}
        \hat{H}_{sp} = \hat{H}_s + \hbar\omega_p \hat{a}_p^\dag \hat{a}_p + i\hbar g_p(\hat{c} \hat{a}_p^\dag - \hat{c}^\dag \hat{a}_p)
    \end{equation}
    gives the coherent exchange of energy between the system and the pseudomode with strength $g_p$, whereas
    \begin{equation}
        \label{eq:LP-coup}
        \hat{H}_{up} = \frac{i\hbar}{2}\sqrt{\gamma_{p}}\big( g_u \hat a_u^\dagger\hat a_{p} - g_u^* \hat a_u\hat a_{p}^\dagger \big)
    \end{equation}
    contains the interaction between the incoming pulse and the pseudomode.
    Dissipation is due to the Lindblad operator
    \begin{equation}
        \label{eq:out-field-u}
        \hat{L}_u = g_u^*\hat{a}_u + \sqrt{\gamma_p}\hat{a}_p.
    \end{equation}
    Note that $\hat{c}$ does not decay directly; it couples to the pseudomode $\hat{a}_p$, which emits at rate $\gamma_p$.

    The Lindblad master equation~\eqref{eq:LMEq-pulse} gives rise to an equivalent Heisenberg equation \cite{CKS17,G77}
    \begin{equation}
        \label{eq:adjoint}
        \begin{split}
            \dot{\hat{c}} & = \frac{i}{\hbar}[\hat{H}_{sp} + \hat{H}_{up},\hat{c}] + \big(\tfrac{\hat{L}_u^\dag}{2} + \hat{b}_{\mathrm{in}}^\dag \big)[\hat{c},\hat{L}_u] + [\hat{L}_u^\dag,\hat{c}]\big(\tfrac{\hat{L}_u}{2} + \hat{b}_{\mathrm{in}}\big),\\
            & = \frac{i}{\hbar}[\hat{H}_s,\hat{c}] + g_p^*[\hat{c}^\dag,\hat{c}]\hat{a}_p.\\
        \end{split}
    \end{equation}
    Indeed, upon inserting the formal solution for $\hat{a}_p(t)$ into Eq.~\eqref{eq:adjoint}, the non-Markovian dynamics~\eqref{eq:eom-sys} is recovered.

    \subsection{Quantum state of the output field}
    \label{ssec:output}

    In general, the quantum state of the scattered light occupies multiple temporal modes and may be highly entangled across them \cite{W89}.
    We are interested in analyzing the state of one of these outgoing modes, $v(t)$.
    Again, this is equivalent to light being absorbed by a virtual cavity, with coupling \cite{KM19}
    \begin{equation}
        \label{eq:in-coup}
        g_v(t)=-\frac{v(t)^*}{\sqrt{\int_{0}^t|v(t^\prime)|^2 dt^\prime}}.
    \end{equation}
    We stress that the time-dependent couplings $g_u(t)$ and $g_v(t)$ are purely theoretical constructs; no such cavities need to be implemented in an experiment~\cite{KM20}.

    Cascading the master equation~\eqref{eq:LMEq-pulse} with the virtual cavity for $v$ \cite{CKS17} collects the part of the multimode field that occupies the mode $\hat{a}_v$.
    When the pulse $u$ scatters off the system into the mode $v$, the state $\hat{\rho}(t)$ of the joint system obeys
    \begin{equation}
        \label{eq:LMEq-pulse-v}
        \dot{\hat{\rho}} = \frac{i}{\hbar}[\hat{\rho},\hat{H}_{sp}+\hat{H}_{upv}] + \hat{L}_0 \hat{\rho} \hat{L}_0^\dag - \tfrac{1}{2}\{\hat{L}_0^\dag \hat{L}_0,\hat{\rho}\}.
    \end{equation}
    where
    \begin{equation}
        \label{eq:LPL-coup}
        \hat{H}_{upv} = \frac{i\hbar}{2}\Big[ \sqrt{\gamma_{p}}g_u \hat a_u^\dagger\hat a_{p} + g_u g_v^*\hat a_u^\dagger \hat a_v + \sqrt{\gamma_{p}}g_v^*\hat a_{p}^\dag\hat a_v - \mathrm{H.c.} \Big]
    \end{equation}
    contains the interaction between the incoming pulse, the pseudomode, and the outgoing light.
    Dissipation is due to the Lindblad operator
    \begin{equation}
        \label{eq:out-field}
        \hat{L}_0 = g_u^*\hat{a}_u + \sqrt{\gamma_p}\hat{a}_p + g_v^*\hat{a}_v.
    \end{equation}

    We stress that the presence of the output cavity does not affect the scattering process.
    This can be verified by deriving the Heisenberg equations for the operators $\hat{a}_u$, $\hat{c}$, and $\hat{a}_p$ which are independent of $\hat{a}_v$; see Sec.~\ref{ssec:markov} below.
    In our construction, it was essential to first construct the pseudomode master equation~\eqref{eq:LMEq-pulse}, which reproduces the exact non-Markovian scattering in Eq.~\eqref{eq:eom-sys}, and only then to cascade the result with the output cavity $v$.
    Otherwise, the virtual cavity of mode $v$ would exert a (nonphysical) back-action on the system.

    \subsection{Markovian limit}
    \label{ssec:markov}

    The Heisenberg equation for the pseudomode is obtained similarly to Eq.~\eqref{eq:adjoint}, viz.,
    \begin{equation}
        \label{eq:eom-pseudo}
        \dot{\hat{a}}_p =  -\big(i\omega_p+\tfrac{\gamma_p}{2}\big)\hat{a}_p + g_p \hat{c}
                     - \sqrt{\gamma_p}\big(g_u^*\hat{a}_u + \hat{b}_{\mathrm{in}}\big).
    \end{equation}
    The Markovian limit is reached when the pseudomode decays much faster than the system evolves, $\gamma_p\to\infty$, while holding the Markovian decay $\gamma = 4g_p^2/\gamma_p$ fixed.
    Any population flowing from the system into the pseudomode is then emitted before it can return.
    Mathematically, this requires an adiabatic elimination of $\hat{a}_p$, i.e., $\dot{\hat{a}}_p\approx 0$, for which Eq.~\eqref{eq:eom-pseudo} implies
    \begin{equation}
        \label{eq:quasistatic}
        \sqrt{\gamma_p}\hat{a}_p
        = \frac{\gamma_p}{i\omega_p+\tfrac{\gamma_p}{2}}
          \Big(\tfrac{\sqrt{\gamma}}{2}\hat{c} + \tilde{g}_u^*\hat{a}_u + \tilde{b}_{\mathrm{in}}\Big).
    \end{equation}
    In the above equation, we applied a phase shift to the coupling, $\tilde{g}_u=-g_u$, and the input field, $\tilde{b}_{\mathrm{in}}=-\hat{b}_{\mathrm{in}}$, as this recovers the convention used in Ref.~\cite{KM19}.
    This has no physical consequence, as Eq.~\eqref{eq:eom-u} remains invariant under this change, i.e., $\tilde{g}_u\tilde{b}_{\mathrm{in}}=g_u\hat{b}_{\mathrm{in}}$.

    For large $\gamma_p$ we have $i\omega_p+\tfrac{\gamma_p}{2}\approx \tfrac{\gamma_p}{2}$ in Eq.~\eqref{eq:quasistatic}, so that the contribution of the (quasistatic) mode $\hat{a}_p$ to Eq.~\eqref{eq:adjoint} yields
    \begin{equation}
        \label{eq:eom-sys-Markov}
        \begin{split}
            \dot{\hat{c}} & \approx \frac{i}{\hbar}[\hat{H}_s,\hat{c}] + [\hat{c}^\dag,\hat{c}]\Big(\tfrac{\gamma}{2}\hat{c} + \sqrt{\gamma}\tilde{g}_u^*\hat{a}_u + \sqrt{\gamma}\tilde{b}_{\mathrm{in}}\Big),
        \end{split}
    \end{equation}
    Finally, the equation of motion for the outgoing mode $v$ is
    \begin{equation}
        \label{eq:eom-out-Markov}
        \begin{split}
            \dot{\hat{a}}_v & = -\frac{|g_v|^2}{2}\hat{a}_v - g_v\big(g_u^*\hat{a}_u + \sqrt{\gamma_p}\hat{a}_p + \hat{b}_{\mathrm{in}}\big)\\
            & \approx -\frac{|g_v|^2}{2}\hat{a}_v - g_v\big(\tilde{g}_u^*\hat{a}_u + \sqrt{\gamma}\hat{c} + \tilde{b}_{\mathrm{in}}\big),
        \end{split}
    \end{equation}
    where we used Eqs.~\eqref{eq:LPL-coup} and~\eqref{eq:quasistatic}.

    The thus obtained equations for $\hat{a}_u$, $\hat{c}$, and $\hat{a}_v$ are equivalent to a master equation with Hamiltonian and Lindblad operator
    \begin{equation}
        \label{eq:KM-recovered}
        \begin{split}
            \hat{H} & \approx \hat{H}_s + \frac{i\hbar}{2}\Big[\sqrt{\gamma}\tilde{g}_u\hat{a}_u^\dag \hat{c}
              + \tilde{g}_u g_v^*\hat{a}_u^\dag\hat{a}_v
              + \sqrt{\gamma} g_v^* \hat{c}^\dag\hat{a}_v
              - \mathrm{H.c.}\Big],\\
            \hat{L}_0 & \approx \tilde{g}_u^*\hat{a}_u + \sqrt{\gamma} \hat{c} + g_v^*\hat{a}_v.
        \end{split}
    \end{equation}
    This recovers the theory of Ref.~\cite{KM19}, as expected.
    Note that, in the Markovian limit, the coherent interaction between the pulse and the system occurs directly through $\hat{c}$ (instead of $\hat{a}_p$).

    \section{Examples}
    \label{sec:apps}

    Here, we apply the theory to the stimulated emission from a two-level atom driven by a single-photon pulse, and to the transmission of a coherent state through an empty cavity.

    \subsection{Stimulated emission by a single-photon pulse}
    \label{ssec:stim-emis}

\begin{figure}[t]
        \centering
        \begin{tikzpicture}
        \node at (0,0) {\includegraphics[width=0.47\textwidth]{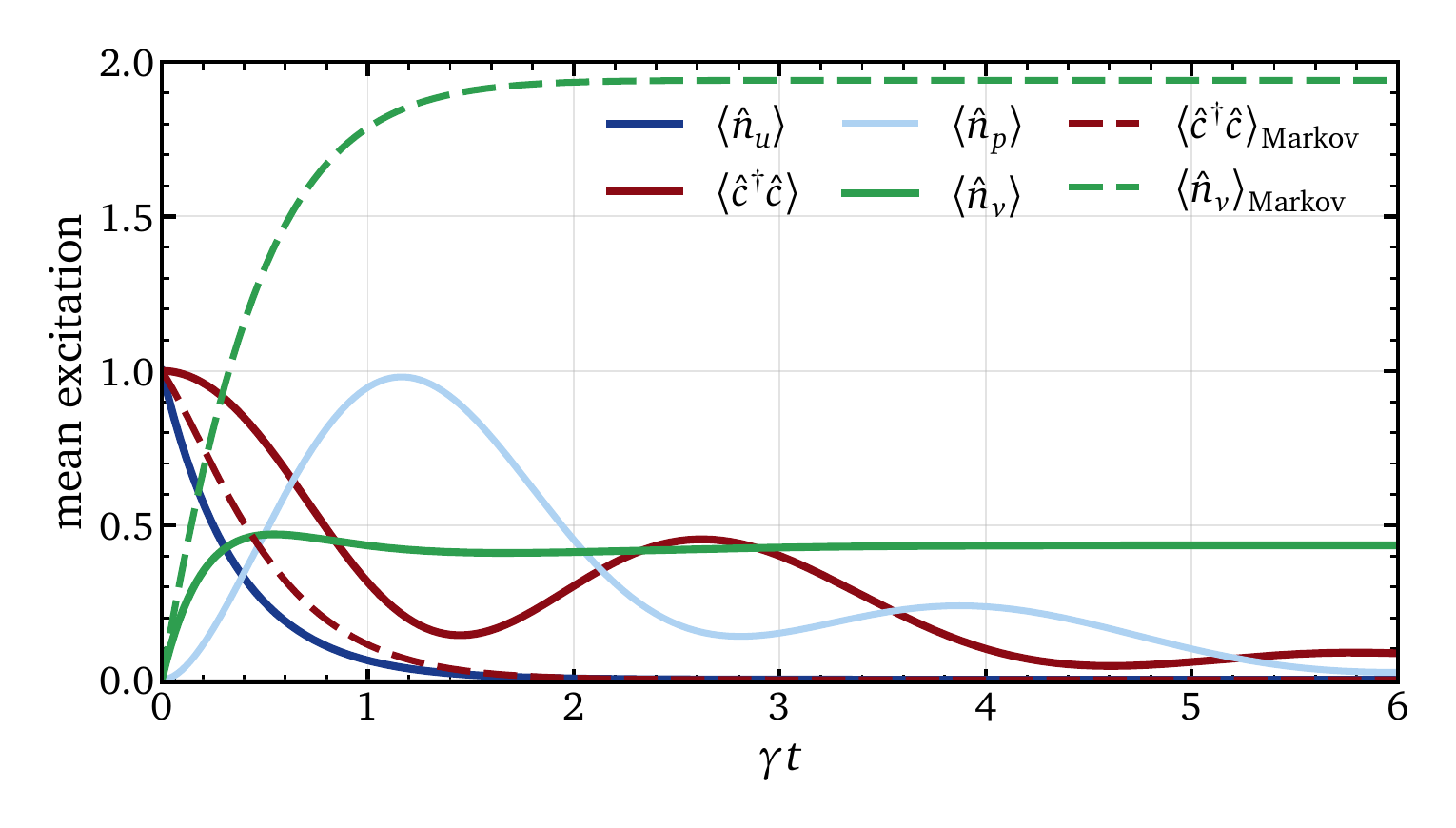}};
        \end{tikzpicture}
        \caption{\label{fig:stim-emis} Stimulated emission from a two-level atom by a single-photon pulse, $\ket{1_u}$.
        Shown are the mean photon numbers $\braket{\hat{n}_k}$ of the incoming pulse ($k=u$), the pseudomode ($k=p$), and the outgoing pulse ($k=v$), together with the atomic population $\braket{\hat{c}^\dag \hat{c}}$.
        Solid lines refer to the non-Markovian dynamics~\eqref{eq:LMEq-pulse-v}, dashed lines to the Markovian dynamics~\eqref{eq:KM-recovered}, for which the population exhibits no revival.
        Parameters used in the simulation are $\Gamma=\gamma/0.36$, $\gamma=\gamma_p$, and $\omega_p=\omega_s=0$.}
    \end{figure}

    Consider a two-level atom that is initially prepared in its excited state $\ket{e}$.
    The atom coherently exchanges energy with a pseudomode $\hat{a}_p$.
    The atom has internal Hamiltonian $\hat{H}_s = \hbar\omega_s \hat{c}^\dag \hat{c}$, with the atomic lowering operator $\hat{c}=\ket{g}\bra{e}$ and resonance frequency $\omega_s$.
    We take the emission to be stimulated by an incoming single-photon pulse, $\ket{1_u}$, in an exponentially decaying mode $u(t)=\sqrt{\Gamma}e^{-\Gamma t/2}$, for $t\geq 0$~\cite{F18}.
    For $\Gamma=\gamma/0.36$, the pulse shape $u(t)$ has been identified as the optimal one for the atom to emit a photon into the outgoing mode $v(t)=u(t)$ \cite{VLP12}.
    In this case, the light-matter couplings in Eqs.~\eqref{eq:out-coup} and \eqref{eq:in-coup} are available in closed form, $g_u(t)=\sqrt{\Gamma}$ and $g_v(t)=\sqrt{\Gamma/(e^{\Gamma t} - 1)}$~\cite{KM19}.
    Figure~\ref{fig:stim-emis} shows that emission into a structured environment falls well short of placing two photons in the desired output mode.
    The non-Markovian revival of the atomic population $\braket{\hat{c}^\dag \hat{c}}$ leads to emission into other modes than $v$, that are of more complex temporal shape.
    This is evident from the fact that the mode $v$ reaches its steady state while the atomic population is still decreasing.
    Hence, we observe a reduced population of the target mode $v$.
    In summary, photon addition is severely hindered by the structured environment, as the light field becomes multimode upon scattering.

    \begin{figure}[t]
        \centering
        \begin{tikzpicture}
        \node at (0,0) {\includegraphics[width=0.47\textwidth]{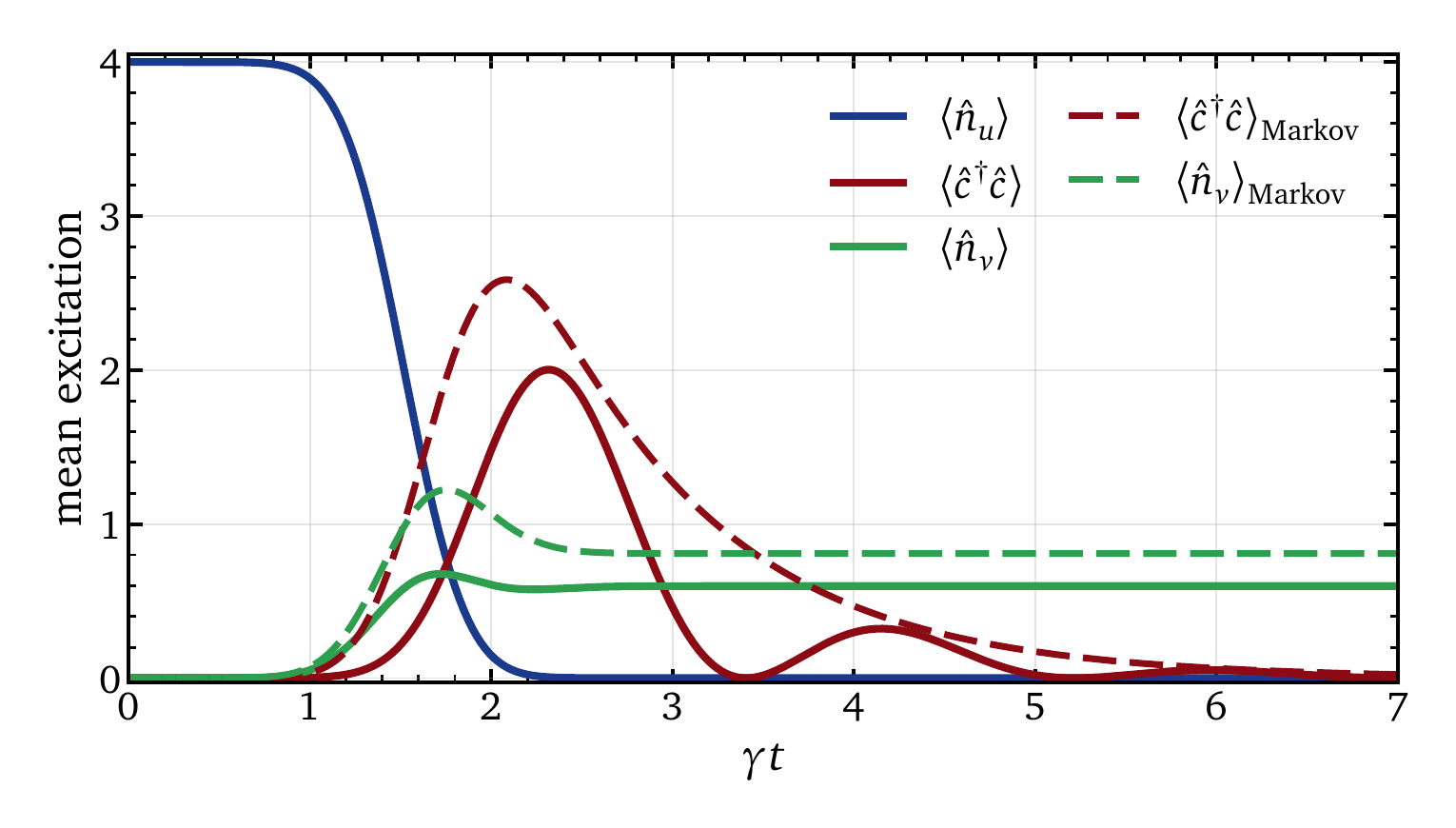}};
        \end{tikzpicture}
        \caption{\label{fig:scat-cav-res} Scattering a Gaussian pulse $u(t)$ off an empty cavity.
        The pulse has width $\tau=0.38/\gamma$ and carries a coherent state $\ket{\alpha_u=2}$.
        Shown are the mean excitations $\braket{\hat{n}_k}$ of the incoming and outgoing pulse ($k=u,v$), together with the intracavity photon number $\braket{\hat{c}^\dag \hat{c}}$.
        Solid lines refer to the non-Markovian evolution, dashed lines to the Markovian evolution.
        The spectral density~\eqref{eq:duo-Lorentz} is specified by $\Omega_{1}=\Omega_{2}=\gamma$, $\gamma_{p_1}=\gamma_{p_2}=\gamma$, and $\omega_{p_1}=-\omega_{p_2}=\gamma$, which corresponds to couplings $g_{p_1}\approx (1.023 - 0.632 i)\gamma$ and $g_{p_2}\approx (0.391 + 0.632 i)\gamma$.}
    \end{figure}

    \subsection{Transmission through an empty cavity}
    \label{ssec:empty-cavity}

    As a second example, we consider the transmission through a cavity with resonance frequency $\omega_c$.
    The internal Hamiltonian of the cavity is $\hat{H}_s = \hbar\omega_c \hat{c}^\dag \hat{c}$, where $\hat{c}$ is the annihilation operator of the cavity mode.
    The cavity is initially in the vacuum state, $\ket{0_s}$.
    We take the coupling of the cavity to the environment to be structured, with a spectral density
    \begin{equation}
        \label{eq:duo-Lorentz}
        \begin{split}
            J(\omega) \propto \frac{\Omega_{1}^2 \gamma_{p_1}}{(\omega - \omega_{p_1})^2 + \big(\frac{\gamma_{p_1}}{2}\big)^2} + \frac{\Omega_{2}^2\gamma_{p_2}}{(\omega - \omega_{p_2})^2 + \big(\frac{\gamma_{p_2}}{2}\big)^2},
        \end{split}
    \end{equation}
    given by two Lorentzians.
    To reproduce the exact reduced dynamics of the light-matter system, two pseudomodes $\hat{a}_{p_1}$ and $\hat{a}_{p_2}$ are required.
    The non-Markovian dynamics are then captured by a Lindblad master equation with Hamiltonian
    \begin{equation}
        \label{eq:Ham-2p}
        \begin{split}
            & \hat{H} = \hat{H}_s + \hbar\omega_{p_1} \hat{a}_{p_1}^\dag\hat{a}_{p_1}
            + \hbar\omega_{p_2} \hat{a}_{p_2}^\dag\hat{a}_{p_2}\\
            & + i\hbar(g_{p_1} \hat{c}\hat{a}_{p_1}^\dag - g_{p_1}^* \hat{c}^\dag\hat{a}_{p_1}) + i\hbar(g_{p_2} \hat{c}\hat{a}_{p_2}^\dag - g_{p_2}^* \hat{c}^\dag\hat{a}_{p_2})\\
            & + \frac{i\hbar}{2}\big(\sqrt{\gamma_{p_1}}g_u\hat{a}_u^\dag \hat{a}_{p_1} + \sqrt{\gamma_{p_2}}g_u\hat{a}_u^\dag\hat{a}_{p_2} + g_ug_v^*\hat{a}_u^\dag \hat{a}_v\\
            &+ \sqrt{\gamma_{p_1}\gamma_{p_2}}\hat{a}_{p_1}^\dag\hat{a}_{p_2} + \sqrt{\gamma_{p_2}}g_v^*\hat{a}_{p_2}^\dag\hat{a}_v + \sqrt{\gamma_{p_1}}g_v^*\hat{a}_{p_1}^\dag\hat{a}_v - \mathrm{H.c.}\big),
        \end{split}
    \end{equation}
    and Lindblad operator
    \begin{equation}
        \label{eq:Lind-2p}
        \hat{L}_0 = g_u^*\hat{a}_u + \sqrt{\gamma_{p_1}}\hat{a}_{p_1} + \sqrt{\gamma_{p_2}}\hat{a}_{p_2} + g_v^*\hat{a}_v,
    \end{equation}
    which are derived in App.~\ref{app:pseudo-Lindblad}.
    There we also show that the couplings $g_{p_1}$ and $g_{p_2}$ to the pseudomodes are not simply given by $\Omega_1$ and $\Omega_2$, but are generically complex.
    This follows from a spectral factorization of $J(\omega)$.

    We consider a Gaussian pulse
    \begin{equation}
        \label{eq:Gauss-pulse}
        u(t) = \frac{1}{\sqrt{\tau} \pi^{1/4}}e^{-\tfrac{(t-t_{c})^2}{2\tau^2}},
    \end{equation}
    that carries a coherent state $\ket{\alpha_u}$.
    Figure~\ref{fig:scat-cav-res} shows the mean photon number of the pulse and inside the cavity, obtained from a solution of the Lindblad master equation with Hamiltonian~\eqref{eq:Ham-2p} and Lindblad operator~\eqref{eq:Lind-2p}.
    The mean photon number $\braket{\hat{c}^\dag \hat{c}}$ inside the cavity displays an oscillatory behavior when the cavity couples to the pseudomodes, which is a signature of the non-Markovian backflow of excitation from the structured environment.

    \section{Conclusion}
    \label{sec:FIN}

    In this article, we generalized the input-output theory for traveling quantum pulses \cite{KM19} to include non-Markovian effects.
    This was achieved by modeling the non-Markovian interaction between a quantum system and its continuum environment in terms of one or several pseudomodes.
    We found that the pseudomode method cannot be applied to the scatterer alone, as doing so assigns each pseudomode its own loss channel, none of which can be associated with the physical output field.
    Instead, the reduced system has to comprise the scatterer together with the incoming pulse.
    It is the non-Markovianity of this joint light-matter system that the pseudomodes reproduce.
    Our approach yields a master equation in Lindblad form for the non-Markovian scattering problem, whose solution is easily obtained using standard computational libraries \cite{JNN12,KPO18}.
    In this master equation, the pseudomodes appeared in a cascaded structure, thereby sidestepping any ambiguity about which pseudomodes have to be identified with the physical output field.

    The theory was applied to the stimulated emission from a two-level atom driven by a single-photon pulse.
    There, we found that emission into a structured environment hinders the creation of a single-mode two-photon pulse.
    Additionally, we studied the transmission of a coherent pulse through an empty cavity whose coupling to the continuum is described by a sum of two Lorentzians.
    In this case, the excitation stored in the cavity oscillates rather than decaying monotonically, due to the backflow of excitation from the environment.
    The advantage of our method is that it incorporates the non-Markovian character of the continuum coupling while still providing the quantum state of the output field.
    Throughout, we have taken the propagation of the pulse to be dispersion free, so that it arrives at the scatterer with the temporal shape with which it was released.
    Note that this is not much of a restriction because any reshaping of the incoming pulse to the dispersion can be absorbed into the definition of the virtual cavity releasing a somewhat deformed pulse onto the scatterer.

    Future work could address the modeling of more complex spectral densities, such as an atom placed inside a material with a photonic band gap \cite{GD06,SFH07}, where the emission is strongly suppressed over a finite frequency window.
    Further natural candidates are systems close to a continuum threshold, such as ionization or dissociation edges, where $J(\omega)$ varies sharply, and regions of strong dispersion, where slow-light effects enhance the light-matter coupling.
    Moreover, the input-output theory is capable of analyzing the quantum state content of more than a single mode of the output field \cite{KM20}.
    Extracting the output eigenmodes would allow one to identify the more exotic temporal shapes that originate from revivals of coherences.
    Finally, the multimode nature of the output state, together with the need for several pseudomodes to model more complex scattering processes, poses a formidable computational challenge owing to the exponential growth of the Fock space of the multimode output state.
    It may be a fruitful endeavor to treat scattering problems using modern computational approaches such as matrix-product states \cite{W92,OR95} and neural quantum states \cite{CT17,TM18,HC19}.

    Our work paves the way for a complete quantum-state description of pulse scattering in structured photonic environments, in which the reshaping of the output field may not only be an imperfection but a degree of freedom to be utilized in the design of light.

    \acknowledgements

    We thank Klaus M\o lmer for helpful comments on the manuscript.

    \appendix

    \section{Linear coupling to a continuum of quantized harmonic oscillators}
    \label{app:lin-coup}

    Here, we state the linear coupling model for a quantum system interacting with a continuum of quantized bosonic field modes \cite{CKS17}.
    The total Hamiltonian,
    \begin{equation}
        \label{aeq:Ham-total}
        \hat{H} = \hat{H}_s + \hat{H}_{I} + \hat{H}_B,
    \end{equation}
    consists of a system Hamiltonian $\hat{H}_s$, the bare Hamiltonian of the bath,
    \begin{equation}
        \label{aeq:Ham-bath}
        \hat{H}_B = \hbar\int_0^\infty \omega \hat{b}^\dag(\omega)\hat{b}(\omega) d\omega,
    \end{equation}
    and the interaction
    \begin{equation}
        \label{aeq:Ham-int}
        \hat{H}_{I} = i\hbar\int_0^\infty \sqrt{J(\omega)}\big[\hat{b}(\omega) + \hat{b}^\dag(\omega)\big]
        \big(e^{i\varphi(\omega)}\hat{c} - e^{-i\varphi(\omega)}\hat{c}^\dag\big) d\omega ,
    \end{equation}
    which describes the linear coupling between the system operator $\hat{c}$ and the bath modes $\hat{b}(\omega)$ at a strength $\sqrt{J(\omega)}$ and phase $\varphi(\omega)$.

    We transform the Hamiltonian $\hat{H}_I$ into the interaction picture with respect to $\hat{H}_s + \hat{H}_B$, viz.,
    \begin{equation}
        \label{aeq:Ham-int-frame}
        \begin{split}
            \tilde{H}_I(t) & = i\hbar\int_0^\infty \sqrt{J(\omega)}\big[e^{i\varphi(\omega)}\hat{c}\hat{b}^\dag(\omega)e^{i(\omega-\Omega)t}\\
            & \quad - e^{-i\varphi(\omega)}\hat{c}^\dag\hat{b}(\omega)e^{-i(\omega-\Omega)t}
            + e^{i\varphi(\omega)}\hat{c}\hat{b}(\omega)e^{-i(\omega+\Omega)t}\\
            & \quad - e^{-i\varphi(\omega)}\hat{c}^\dag\hat{b}^\dag(\omega)e^{i(\omega+\Omega)t}\big]d\omega.\\
        \end{split}
    \end{equation}
    Here, we have taken the system operator to satisfy $\tilde{c}=e^{i \hat{H}_s t/\hbar}\hat{c}e^{-i \hat{H}_s t/\hbar}=\hat{c}e^{-i\Omega t}$, for some frequency $\Omega>0$.
    The rotating-wave approximation drops the terms oscillating with $\omega+\Omega$.
    Doing so and transforming back into the lab frame gives
    \begin{equation}
        \label{aeq:RWA-int}
        \hat{H}_{I}\approx i\hbar\int_0^\infty \sqrt{J(\omega)}\big[e^{i\varphi(\omega)}\hat{c}\hat{b}^\dag(\omega) - \mathrm{H.c.}\big]d\omega,
    \end{equation}
    which is the interaction stated in Sec.~\ref{sec:theory}.

    \section{Heisenberg equations for the scattering process}
    \label{app:Langevin}

    Here, we derive the Heisenberg equations for the scattering process.
    That is, a pulse $u(t)$ couples to the continuum and scatters off a quantum system that emits into a structured environment.
    Assuming a coupling as in Eq.~\eqref{aeq:RWA-int}, the full Hamiltonian is given by
    \begin{equation}
        \label{aeq:full-Ham}
        \begin{split}
            \hat{H} & = \hat{H}_s + \hat{H}_B + i\hbar\int_{0}^\infty \Big(G_u(\omega,t)\hat{a}_u\hat{b}^\dag(\omega)\\
            & \quad + \sqrt{J(\omega)}e^{i(\varphi(\omega)-\omega\tau_c)}\hat{c}\hat{b}^\dag(\omega) - \mathrm{H.c.}\Big)d\omega,\\
        \end{split}
    \end{equation}
    with a spectral density $J(\omega)$ and $G_u(\omega,t)=g_u^*(t)e^{-i\omega\tau_u}/\sqrt{2\pi}$, where $g_u$ is specified in Eq.~\eqref{eq:out-coup}.
    The phases $e^{-i\omega\tau_u}$ and $e^{-i\omega\tau_c}$ account for the propagation delays, $\tau_u<\tau_c$.

    The Heisenberg equation for the continuum mode is
    \begin{equation}
        \label{aeq:eom-bath}
        \begin{split}
            \dot{\hat{b}}(\omega) & = \frac{i}{\hbar}[\hat{H},\hat{b}(\omega)]\\
            & = -i\omega\hat{b}(\omega) + G_u \hat{a}_u + \sqrt{J(\omega)}e^{i(\varphi(\omega)-\omega\tau_c)}\hat{c},
        \end{split}
    \end{equation}
    where we used $[\hat{b}(\omega),\hat{b}^\dag(\omega^\prime)]=\delta(\omega-\omega^\prime)$.
    Equation~\eqref{aeq:eom-bath} has the formal solution
    \begin{equation}
    \label{aeq:formal-sol}
    \begin{split}
        \hat{b}(\omega,t) & = e^{-i\omega t}\hat{b}(\omega,0)
        + \int_0^t e^{-i\omega(t-t^\prime)}\Big(G_u(\omega,t^\prime)\hat{a}_u(t^\prime)\\
        & \quad + \sqrt{J(\omega)}e^{i(\varphi(\omega)-\omega\tau_c)}\hat{c}(t^\prime)\Big)dt^\prime ,
    \end{split}
    \end{equation}
    which, inserted into the equation $\hat{a}_u$, gives
    \begin{equation}
        \label{aeq:eom-u-mode}
        \begin{split}
            \dot{\hat{a}}_u & = \frac{i}{\hbar}[\hat{H},\hat{a}_u] = -\int_0^\infty G_u^*(\omega,t)\hat{b}(\omega,t)d\omega\\
            & = -g_u(t)\Big[\hat{b}_{\mathrm{in}}(t-\tau_u) + \int_0^t \delta(t-t^\prime)
            g_u^*(t^\prime)\hat{a}_u(t^\prime)dt^\prime\\
            &\quad + \int_0^t K^*\big(t^\prime-t+\tau_c-\tau_u\big)
            \hat{c}(t^\prime)dt^\prime\Big].
        \end{split}
    \end{equation}
    Here we introduced the input field
    \begin{equation}
        \label{aeq:input-field}
        \hat{b}_{\mathrm{in}}(t) = \frac{1}{\sqrt{2\pi}}\int_0^\infty
        e^{-i\omega t}\hat{b}(\omega,0)d\omega,
    \end{equation}
    and the kernel
    \begin{equation}
        \label{aeq:amp-ker}
        K(t-t^\prime) = \frac{1}{\sqrt{2\pi}}\int_0^\infty \sqrt{J(\omega)} e^{-i\varphi(\omega)}e^{-i\omega (t-t^\prime)}d\omega.
    \end{equation}
    In carrying out the $\omega$-integration we extended the lower limit of the frequency integral to $-\infty$, which is legitimate as long as $J(\omega)$ is centered at an optical frequency, and then used $\delta(t)=\frac{1}{2\pi}\int_{-\infty}^\infty e^{i\omega t} d\omega$.
    To evaluate each integral in Eq.~\eqref{aeq:eom-u-mode} we make use of the causal ordering $\tau_u<\tau_c$ of the scattering.
    Regularizing $\delta(t)$ to $\tfrac{\lambda}{2}e^{-\lambda|t|}$ in the limit of $\lambda\to\infty$, we find
    \begin{equation}
        \label{aeq:half-weight}
        \int_0^t \delta(t-t^\prime)g_u^*(t^\prime)\hat{a}_u(t^\prime)dt^\prime=\frac{g_u^*(t)}{2}\hat{a}_u(t).
    \end{equation}
    The integral over $K$ is non-zero only for $t^\prime-t+\tau_c-\tau_u\geq 0$, or $t^\prime\geq t-(\tau_c-\tau_u)$.
    This lies beyond the upper limit of the integral, so that the integral vanishes.
    Taking the time delay to be negligible, $\tau_u\to 0$, the Heisenberg equation for $\hat{a}_u$ simplifies to
    \begin{equation}
        \label{aeq:mode-u-final}
        \dot{\hat{a}}_u = - \frac{1}{2}|g_u(t)|^2\hat{a}_u - g_u(t)\hat{b}_{\mathrm{in}}(t).
    \end{equation}
    The equation for $\hat{c}$ is derived analogously, leading to
    \begin{equation}
        \label{aeq:eom-sys}
        \begin{split}
            \dot{\hat{c}} & = \frac{i}{\hbar}[\hat{H}_s,\hat{c}] + [\hat{c}^\dag,\hat{c}]\bigg(\hat{f}_{\mathrm{in}}(t) + \int_0^t C(t-t^\prime)\hat{c}(t^\prime)dt^\prime\\
            & \quad + \int_0^t K(t-t^\prime)g_u^*(t^\prime)\hat{a}_u(t^\prime)dt^\prime\bigg).\\
        \end{split}
    \end{equation}
    Here, we defined the memory kernel
    \begin{equation}
        \label{aeq:bath-corr}
        C(t-t^\prime) = \int_0^\infty J(\omega) e^{-i\omega(t-t^\prime)} d\omega,
    \end{equation}
    as well as the colored input field
    \begin{equation}
    \label{aeq:fin}
    \hat{f}_{\mathrm{in}}(t) = \int_0^t K(t-t^\prime)
        \hat{b}_{\mathrm{in}}(t^\prime)dt^\prime,
    \end{equation}
    i.e., $[\hat{f}_{\mathrm{in}}(t),\hat{f}_{\mathrm{in}}^\dag(t^\prime)]=C(t-t^\prime)$.

    \section{Lorentzian-sum model of the spectral density}
    \label{app:sum-Lorentz}

    Consider the spectral density
    \begin{equation}
        \label{aeq:sum-Lorentz}
        J(\omega) = \sum_{k=1}^N \frac{\Omega_k^2}{2\pi} \frac{\gamma_k}{(\omega-\omega_k)^2 + \left(\frac{\gamma_k}{2}\right)^2},
    \end{equation}
    which is a finite sum of $N$ Lorentzian functions with real-valued $\Omega_k$, $\gamma_k$, and $\omega_k$.
    The corresponding kernel~\eqref{aeq:bath-corr} is obtained via Fourier transform,
    \begin{equation}
        \label{aeq:bath-Lorentz}
        \begin{split}
            C(t-t^\prime) & \approx \int_{-\infty}^\infty J(\omega) e^{-i\omega(t-t^\prime)} d\omega\\
            & = \sum_k \Omega_k^2 e^{-(i\omega_k + \gamma_k/2)(t-t^\prime)},
        \end{split}
    \end{equation}
    where the extension of the lower limit of the integral to $-\infty$ is valid for $\gamma_k \ll \omega_k$.

    To obtain the amplitude kernel in Eq.~\eqref{aeq:amp-ker} we have to determine $\sqrt{J(\omega)}e^{-i\varphi(\omega)}$.
    Write Eq.~\eqref{aeq:sum-Lorentz} over a common denominator,
    \begin{equation}
        \label{aeq:J-rational}
        \begin{split}
            J(\omega) & = \sum_{k} \frac{\Omega_k^2}{2\pi} \frac{\gamma_k}{(\omega-\tilde{\omega}_k)(\omega-\tilde{\omega}_k^*)},\\
            & = \frac{P(\omega)}{\prod_{j}(\omega-\tilde{\omega}_j)(\omega-\tilde{\omega}_j^*)},\\
        \end{split}
    \end{equation}
    with poles $\tilde{\omega}_j = \omega_j - i\gamma_j/2$ and the polynomial
    \begin{equation}
        \label{aeq:numerator}
        P(\omega) = \frac{1}{2\pi}\sum_k \Omega_k^2\gamma_k
        \prod_{j\neq k}(\omega-\tilde{\omega}_j)(\omega-\tilde{\omega}_j^*).
    \end{equation}
    Since $\omega$ is real, each term in Eq.~\eqref{aeq:numerator} equals
    \begin{equation}
        \label{aeq:positivity}
        \frac{\Omega_k^2\gamma_k}{2\pi}\prod_{j\neq k}|\omega-\tilde{\omega}_j|^2>0.
    \end{equation}
    Then, $P$ is a polynomial of degree $2(N-1)$ with real coefficients that is strictly positive on the real axis.
    Its roots therefore form complex-conjugate pairs $(z_m,z_m^*)_{m=1}^{N-1}$ with no real elements.
    Choosing these in the lower half-plane, $\mathrm{Im}(z_m)<0$, fixes the phase factor $\varphi(\omega)$, i.e.,
    \begin{equation}
        \label{aeq:eta-factor}
        \sqrt{J(\omega)}e^{-i\varphi(\omega)} = -i\sqrt{\beta}
        \frac{\prod_{m=1}^{N-1}(\omega-z_m)}{\prod_{j=1}^{N}(\omega-\tilde{\omega}_j)},
    \end{equation}
    with $\beta = \sum_k \Omega_k^2\gamma_k/(2\pi)$
    One readily verifies that, for real $\omega$, the above equation satisfies $|\sqrt{J(\omega)}e^{-i\varphi(\omega)}|^2 = J(\omega)$ as intended.
    The choice of phase in Eq.~\eqref{aeq:eta-factor} ensures that $K(s)=0$ for $s<0$, as required in Eq.~\eqref{aeq:eom-u-mode}.
    Since all poles of Eq.~\eqref{aeq:eta-factor} coincide with those of $J(\omega)$ in the lower half-plane, both kernels share their exponents,
    \begin{equation}
        \label{aeq:K-Lorentz}
        \begin{split}
            K(t-t^\prime) & = \sum_j A_je^{-(i\omega_j+\gamma_j/2)(t-t^\prime)},\\
            A_j & = -\sqrt{2\pi\beta}\frac{\prod_{m}(\tilde{\omega}_j-z_m)}{\prod_{l\neq j}(\tilde{\omega}_j-\tilde{\omega}_l)},
        \end{split}
    \end{equation}
    only the $A_j$ differing from the $\Omega_j^2$ in Eq.~\eqref{aeq:bath-Lorentz}.
    We stress that the factorization~\eqref{aeq:eta-factor} is not unique, due to freedom in the choice of $z_m$.
    The reduced dynamics of the scatterer is insensitive to this choice.

    Clearly, the memory kernels, $C\propto J(\omega)$ and $K\propto\sqrt{J(\omega)}$, are not independent.
    To determine how they are related we use
    \begin{equation}
        \int_0^\infty K(s+\tau)K^*(s)ds = C(\tau),
    \end{equation}
    which follows from the definition of $\hat{f}_{\mathrm{in}}$ in Eq.~\eqref{aeq:fin}.
    Evaluating this integral for the kernel~\eqref{aeq:K-Lorentz} brings us to
    \begin{equation*}
        \begin{split}
            \int_0^\infty K(s+\tau)K^*(s)ds & = \sum_{j,k} A_j A_k^* e^{-\Lambda_j \tau}\int_0^\infty e^{-(\Lambda_j + \Lambda_k^*)s} ds,\\
            & = \sum_{j,k} \frac{A_j A_k^*}{\Lambda_j + \Lambda_k^*} e^{-\Lambda_j \tau},\\
        \end{split}
    \end{equation*}
    where we defined $\Lambda_j = i\omega_{j}+\gamma_{j}/2$.
    Comparing with the coefficients of Eq.~\eqref{aeq:bath-Lorentz} yields the system of equations
    \begin{equation}
        \label{aeq:autocorr}
            \sum_{k}\frac{A_j A_k^*}{i(\omega_j-\omega_k)+(\gamma_j+\gamma_k)/2} = \Omega_j^2 ,
    \end{equation}
    for every $j$.
    This provides a recipe to relate the constants $A_j$ to the parameters $\Omega_j$.

    \subsection{Special cases}
    \label{app:examples}

    Consider a single Lorentzian, $N=1$.
    Then, Eq.~\eqref{aeq:autocorr} gives immediately $|A_1|=\Omega_1\sqrt{\gamma_1}$.
    With Eq.~\eqref{aeq:K-Lorentz} and the definition of $\beta$ we get $A_1=-\Omega_1\sqrt{\gamma_1}$.
    For $N=2$, we have $P(\omega)=\beta(\omega^2-2x_1\omega+x_2)$ with
    \begin{equation}
        \label{aeq:N2-coeffs}
        \begin{split}
            x_1 & = \frac{\Omega_1^2\gamma_1\omega_2+\Omega_2^2\gamma_2\omega_1}
                 {\Omega_1^2\gamma_1+\Omega_2^2\gamma_2},\\
        x_2 & = \frac{\Omega_1^2\gamma_1\big(\omega_2^2+\gamma_2^2/4\big)
                   + \Omega_2^2\gamma_2\big(\omega_1^2+\gamma_1^2/4\big)}
                 {\Omega_1^2\gamma_1+\Omega_2^2\gamma_2},
        \end{split}
    \end{equation}
    so that $z_1=x_1 - i\sqrt{x_2-x_1^2}$ and
    \begin{equation}
        \label{aeq:N2-A}
        A_1 = -\sqrt{2\pi\beta}\frac{\tilde{\omega}_1-z_1}{\tilde{\omega}_1-\tilde{\omega}_2},
        \quad
        A_2 = -\sqrt{2\pi\beta}\frac{\tilde{\omega}_2-z_1}{\tilde{\omega}_2-\tilde{\omega}_1}.
    \end{equation}

    \section{Pseudomode approach to quantum scattering}
    \label{app:pseudo-Lindblad}

    Here, we show that the reduced dynamics in Eq.~\eqref{aeq:eom-sys} for the Lorentzian-sum model~\eqref{aeq:sum-Lorentz} are equivalently described by a collection of cascaded pseudomodes $\hat{a}_{p_k}$.
    The full system-reservoir Hamiltonian for this model is
    \begin{equation}
        \label{aeq:pseudo-Ham}
        \begin{split}
            \hat{H} & = \hat{H}_s + \hbar\sum_k \omega_{p_k} \hat{a}_{p_k}^\dag \hat{a}_{p_k}
                + i\hbar\sum_k \big(g_{p_k} \hat{c}\hat{a}_{p_k}^\dag - g_{p_k}^* \hat{c}^\dag \hat{a}_{p_k}\big)\\
              & \quad + i\hbar\int_0^\infty \Big(G_u(\omega,t)\hat{a}_u \hat{b}^\dag(\omega)
                + G_v(\omega,t)\hat{a}_v \hat{b}^\dag(\omega)\\
              &\quad + \sum_k G_{p_k}(\omega)\hat{a}_{p_k}\hat{b}^\dag(\omega) - \mathrm{H.c.}\Big)d\omega  + \hat{H}_B,
        \end{split}
    \end{equation}
    where $G_k(\omega,t) = g_k^*(t)e^{-i\omega \tau_k}/\sqrt{2\pi}$ for $k=u,v$ and the couplings $g_u$ and $g_v$ are given in Eqs.~\eqref{eq:out-coup} and \eqref{eq:in-coup}, respectively.
    Note that the scatterer no longer couples to the continuum.
    Instead, it is coupled coherently, with strengths $g_{p_k}$, to a collection of pseudomodes whose
    couplings to the continuum are flat, $G_{p_k}(\omega) = \sqrt{\gamma_{p_k}/2\pi}e^{-i\omega\tau_{p_k}}$.
    The propagation delays are ordered as
    \begin{equation}
        \label{aeq:delays}
        \tau_u < \tau_{p_1} < \dots < \tau_{p_N} < \tau_v.
    \end{equation}

    Inserting the formal solution for the continuum mode $\hat{b}(\omega,t)$ of the model~\eqref{aeq:pseudo-Ham}, the Heisenberg equations for the relevant operators read
    \begin{equation}
        \label{aeq:pseudo-all}
        \begin{split}
            \dot{\hat{a}}_u & = - \frac{1}{2}\big|g_u\big|^2 \hat{a}_u - g_u\hat{b}_{\mathrm{in}},\\
            \dot{\hat{a}}_{p_k} & = -\bigg(i\omega_{p_k}+\frac{\gamma_{p_k}}{2}\bigg)\hat{a}_{p_k} -\sum_{j<k}\sqrt{\gamma_{p_j}\gamma_{p_k}}\hat{a}_{p_j} + g_{p_k}\hat{c} \\
            &\quad - \sqrt{\gamma_{p_k}}\Big(g_u^*\hat{a}_u + \hat{b}_{\mathrm{in}}\Big),\\
            \dot{\hat{c}} & = \frac{i}{\hbar}[\hat{H}_s,\hat{c}] + [\hat{c}^\dag,\hat{c}]\sum_k g_{p_k}^*\hat{a}_{p_k},\\
            \dot{\hat{a}}_v & = - \frac{1}{2}\big|g_v\big|^2 \hat{a}_v - g_v\Big(g_u^*\hat{a}_u + \sum_k\sqrt{\gamma_{p_k}}\hat{a}_{p_k} + \hat{b}_{\mathrm{in}}\Big).
        \end{split}
    \end{equation}
    No explicit memory kernel is present in the above equations, as each pseudomode couples to a Markovian reservoir.
    Note further that $\hat{a}_v$ appears only in its own equation, so that the output cavity exerts no back-action on $\hat{a}_u$, $\hat{a}_{p_k}$, and $\hat{c}$.

    It is convenient to collect the pseudomodes into a vector
    $\hat{\bm{a}}_p = (\hat{a}_{p_1},\dots,\hat{a}_{p_N})^T$, so that the second line of
    Eq.~\eqref{aeq:pseudo-all} reads
    \begin{equation}
        \label{aeq:pseudo-vector}
        \dot{\hat{\bm{a}}}_p = -\bm{\Gamma}\hat{\bm{a}}_p + \bm{g}_p \hat{c} - \bm{\gamma}\big(g_u^*\hat{a}_u + \hat{b}_{\mathrm{in}}\big),
    \end{equation}
    with $\bm{g}_p = (g_{p_1},\dots,g_{p_N})^T$, $\bm{\gamma} = (\sqrt{\gamma_{p_1}},\dots,\sqrt{\gamma_{p_N}})^T$, and the lower
    triangular matrix
    \begin{equation}
        \label{aeq:Gamma-matrix}
        (\bm{\Gamma})_{kj} =
        \begin{cases}
            \Lambda_k, & j = k,\\
            \sqrt{\gamma_{p_k}\gamma_{p_j}}, & j<k,\\
            0, & j>k,
        \end{cases}
        \qquad \Lambda_k = i\omega_{p_k} + \frac{\gamma_{p_k}}{2}.
    \end{equation}
    The formal solution of Eq.~\eqref{aeq:pseudo-vector} is then
    \begin{equation}
        \label{aeq:pseudo-matrix-sol}
        \hat{\bm{a}}_p(t) = \int_0^t e^{-\bm{\Gamma}(t-t^\prime)}
        \big[\bm{g}_p \hat{c}(t^\prime) - \bm{\gamma}\big(g_u^*\hat{a}_u(t^\prime) + \hat{b}_{\mathrm{in}}(t^\prime)\big)\big]dt^\prime ,
    \end{equation}
    with each pseudomode being initially in the vacuum state.
    The non-Markovian nature of the scatterer is recovered upon inserting the pseudomodes,
    \begin{equation}
    \label{aeq:eliminated-braced}
    \begin{split}
        \dot{\hat{c}} & = \frac{i}{\hbar}[\hat{H}_s,\hat{c}] + [\hat{c}^\dag,\hat{c}]\int_0^t\bigg[(\bm{g}_p^*)^T e^{-\bm{\Gamma}(t-t^\prime)}\bm{g}_p \hat{c}(t^\prime)\\
        & \quad - (\bm{g}_p^*)^T e^{-\bm{\Gamma}(t-t^\prime)}\bm{\gamma}\big(g_u^*\hat{a}_u(t^\prime) + \hat{b}_{\mathrm{in}}(t^\prime)\big)\bigg]dt^\prime.
    \end{split}
    \end{equation}

    The aim of the pseudomode approach is to reproduce the exact non-Markovian dynamics of the scatterer in Eq.~\eqref{aeq:eom-sys}.
    Comparing Eq.~\eqref{aeq:eliminated-braced} with Eq.~\eqref{aeq:eom-sys} identifies the memory kernels as
    \begin{equation}
        \label{aeq:chain-kernels}
        \begin{split}
            C(\tau) & = (\bm{g}_p^*)^T e^{-\bm{\Gamma}\tau}\bm{g}_p = \sum_k C_k e^{-(i\omega_{p_k}+\gamma_{p_k}/2)\tau},\\
            K(\tau) & = - (\bm{g}_p^*)^T e^{-\bm{\Gamma}\tau}\bm{\gamma} = \sum_k K_k e^{-(i\omega_{p_k}+\gamma_{p_k}/2)\tau}.\\
        \end{split}
    \end{equation}
    The coefficients $C_k$ and $K_k$ are precisely the poles of the Laplace transformed kernels,
    \begin{equation}
        \label{aeq:chain-laplace}
        \begin{split}
            \tilde{C}(s) & = (\bm{g}_p^*)^T\big(s\mathbb{I}+\bm{\Gamma}\big)^{-1}\bm{g}_p,\\
            \tilde{K}(s) & = -(\bm{g}_p^*)^T\big(s\mathbb{I}+\bm{\Gamma}\big)^{-1}\bm{\gamma},\\
        \end{split}
    \end{equation}
    where $\mathbb{I}$ is the $N\times N$ unit matrix.
    Since $\bm{\Gamma}$ is triangular, its eigenvalues are its diagonal entries, so that $\tilde{C}$ and $\tilde{K}$ have simple poles at $-\Lambda_k$ only.
    To match the Heisenberg equation~\eqref{aeq:eliminated-braced} onto Eq.~\eqref{aeq:eom-sys} it suffices to set $K_k=A_k$, with $A_k$ defined in Eq.~\eqref{aeq:K-Lorentz}.

    \subsection{Special cases}

    To establish the equivalence of Eq.~\eqref{aeq:eliminated-braced} with Eq.~\eqref{aeq:eom-sys}, we clearly need $\gamma_{p_k}=\gamma_k$ and $\omega_{p_k}=\omega_k$ for the spectral density in Eq.~\eqref{aeq:sum-Lorentz}.
    The couplings $g_{p_k}$ must then be chosen such that the kernel $K$ reproduces Eq.~\eqref{aeq:K-Lorentz}.
    Though, this can be done in general using Eq.~\eqref{aeq:chain-laplace}, it is more instructive to illustrate the strategy on simple examples.

    For a single Lorentzian, $N=1$, we have $\tilde{K}(s)=-g_{p_1}^*\sqrt{\gamma_{p_1}}/(s+\Lambda_1)$ from Eq.~\eqref{aeq:chain-laplace}.
    Since $s=-\Lambda_1$ is a simple pole we get
    \begin{equation}
        K_1 = \lim\limits_{s\to -\Lambda_1} (s+\Lambda_1)\tilde{K}(s) = -g_{p_1}^*\sqrt{\gamma_{p_1}}.
    \end{equation}
    When matched to $A_1=-\Omega_1\sqrt{\gamma_1}$ from App.~\ref{app:examples} we find $g_{p_1}=\Omega_1$.

    For $N=2$, the same procedure gives
    \begin{equation}
        \label{aeq:forward-N2}
            \begin{split}
            A_1 & = -g_{p_1}^*\sqrt{\gamma_{p_1}}
              +\frac{\gamma_{p_1}g_{p_2}^*\sqrt{\gamma_{p_2}}}{\Lambda_2-\Lambda_1},\\
            A_2 & = -g_{p_2}^*\sqrt{\gamma_{p_2}}\frac{\Lambda_1^*+\Lambda_2}{\Lambda_2-\Lambda_1}.
            \end{split}
    \end{equation}
    Basic algebra yields the couplings as
    \begin{equation}
        \label{aeq:inverse-N2}
        \begin{split}
        g_{p_1} & = -\frac{1}{\sqrt{\gamma_{p_1}}}
          \left[A_1^* + \frac{\gamma_{p_1}A_2^*}{\Lambda_1+\Lambda_2^*}\right],\\
        g_{p_2} & = -\frac{A_2^*}{\sqrt{\gamma_{p_2}}}
          \frac{\Lambda_2^*-\Lambda_1^*}{\Lambda_1+\Lambda_2^*}.
        \end{split}
    \end{equation}
    Note that the $g_{p_k}$ are generically complex, even though the $\Omega_k$ in Eq.~\eqref{aeq:sum-Lorentz} are real.

    For $\Omega_k=\gamma_{p_k}=\gamma$ and $\omega_{p_{1,2}}=\pm \gamma$, we obtain
    $g_{p_1}\approx (1.023 - 0.632 i)\gamma$ and $g_{p_2}\approx (0.391 + 0.632 i)\gamma$.
    Equivalently, we have
    \begin{equation}
        \label{aeq:gp-invariants}
        |g_{p_{1,2}}|^2 = \Big(1\pm\frac{1}{\sqrt{5}}\Big)\gamma^2,\quad  |g_{p_{1}}|^2 + |g_{p_{2}}|^2 = \Omega_{1}^2 + \Omega_{2}^2 .
    \end{equation}
    This fixes the parameters used in the calculation of the mean excitation shown in Fig.~\ref{fig:scat-cav-res}.

    \subsection{Pseudomode master equation}

    It remains to be shown that the master equation
    \begin{equation}
        \label{aeq:pseudo-master}
        \dot{\hat{\rho}} = \frac{i}{\hbar}[\hat{\rho},\hat{H}_{sp} + \hat{H}_{upv}] + \hat{L}_0\hat{\rho} \hat{L}_0^\dag
          - \tfrac{1}{2}\big\{\hat{L}_0^\dag \hat{L}_0,\hat{\rho}\big\},
    \end{equation}
    with Hamiltonian
    \begin{equation}
        \label{aeq:Hsys}
            \hat{H}_{sp} = \hat{H}_s + \hbar\sum_k \omega_{p_k} \hat{a}_{p_k}^\dag\hat{a}_{p_k}
          + i\hbar\sum_k \big(g_{p_k} \hat{c}\hat{a}_{p_k}^\dag
            - g_{p_k}^* \hat{c}^\dag\hat{a}_{p_k}\big),
    \end{equation}
    and Lindblad operator
    \begin{equation}
        \label{aeq:collapse}
        \hat{L}_0(t) = g_u^*(t)\hat{a}_u + \sum_k \sqrt{\gamma_{p_k}}\hat{a}_{p_k}
                   + g_v^*(t)\hat{a}_v,
    \end{equation}
    is equivalent to the Heisenberg equations \eqref{aeq:pseudo-all}.
    In Eq.~\eqref{aeq:pseudo-master}, the interaction between the coherent interaction between the pulse and the pseudomodes is due to
    \begin{equation}
        \label{aeq:Hcasc}
        \hat{H}_{upv}(t) = \frac{i\hbar}{2}\sum_{a<b}
          \big[\hat{X}_a^\dag(t) \hat{X}_b(t) - \hat{X}_a(t)\hat{X}_b^\dag(t)\big],
    \end{equation}
    where $\hat{X}_u = g_u^*\hat{a}_u$, $\hat{X}_{p_k}= \sqrt{\gamma_{p_k}}\hat{a}_{p_k}$, and
    $\hat{X}_v = g_v^*\hat{a}_v$, ordered by their delays~\eqref{aeq:delays}.
    To show the equivalence to Eq.~\eqref{aeq:pseudo-all}, recall that a Lindblad master equation as in Eq.~\eqref{aeq:pseudo-master} is connected to a Heisenberg equation via \cite{CKS17}
    \begin{equation}
        \label{aeq:adjoint}
        \begin{split}
            \dot{\hat{X}} & = \frac{i}{\hbar}[\hat{H},\hat{X}] + \Big(\tfrac{\hat{L}_0^\dag}{2} + \hat{b}_{\mathrm{in}}^\dag\Big)[\hat{X},\hat{L}_0] + [\hat{L}_0^\dag,\hat{X}]\Big(\tfrac{\hat{L}_0}{2} + \hat{b}_{\mathrm{in}}\Big)\\
        \end{split}
    \end{equation}
    for any operator $\hat{X}$, and $\hat{H}=\hat{H}_{sp}+\hat{H}_{upv}$.
    The desired equivalence follows now from a straightforward calculation.
    For a single pseudomode, $N=1$, we recover the case discussed in Sec.~\ref{sec:theory}.
    For two pseudomodes, $N=2$, we find the equations studied in Sec.~\ref{ssec:empty-cavity}.

\end{document}